# Frankfurt Big Data Lab
Databases and Information Systems (DBIS)

GOETHE UNIVERSITÄT
FRANKFURT AM MAIN

# Evaluating Hadoop Clusters with TPCx-HS

*Technical Report No. 2015-1*
*October 6, 2015*

Todor Ivanov and Sead Izberovic

Frankfurt Big Data Lab
Chair for Databases and Information Systems
Institute for Informatics and Mathematics
Goethe University Frankfurt
Robert-Mayer-Str. 10,
60325, Bockenheim
Frankfurt am Main, Germany

*www.bigdata.uni-frankfurt.de*

# Table of Contents

# 1. Introduction

The growing complexity and variety of Big Data platforms makes it both difficult and time consuming for all system users to properly setup and operate the systems. Another challenge is to compare the platforms in order to choose the most appropriate one for a particular application. All these factors motivate the need for a standardized Big Data benchmark that can help the users in the process of platform evaluation. Just recently TPCx-HS [1][2] has been released as the first standardized Big Data benchmark designed to stress test a Hadoop cluster.

The goal of this study is to evaluate and compare how the network setup influences the performance of a Hadoop cluster. In particular, experiments were performed using shared and dedicated 1Gbit networks utilized by the same Cloudera Hadoop Distribution (CDH) cluster setup. The TPCx-HS benchmark, which is very network intensive, was used to stress test and compare both cluster setups. All the presented results are obtained by using the officially available version [1] of the benchmark, but they are not comparable with the officially reported results and are meant as an experimental evaluation, not audited by any external organization. As expected the dedicated 1Gbit network setup performed much faster than the shared 1Gbit setup. However, what was surprising is the negligible price difference between both cluster setups, which pays off with a multifold performance return.

The rest of the report is structured as follows: Section 2 provides a brief description of the technologies involved in our study. An overview of the hardware and software setup used for the experiments is given in Section 3. Brief summary of the TPCx-HS benchmark is presented in Section 4. The performed experiments together with the evaluation of the results are presented in Section 5. Finally, Section 6 concludes with lessons learned.

# 2. Background

**Big Data** has emerged as a new term not only in IT, but also in numerous other industries such as healthcare, manufacturing, transportation, retail and public sector administration [3][4] where it quickly became relevant. There is still no single definition which adequately describes all Big Data aspects [5], but the "*V*" characteristics (*Volume*, *Variety*, *Velocity*, *Veracity* and more) are among the widely used one. Exactly these new Big Data characteristics challenge the capabilities of the traditional data management and analytical systems [5][6]. These challenges also motivate the researchers and industry to develop new types of systems such as Hadoop and NoSQL databases [7].

**Apache Hadoop** [8] is a software framework for distributed storing and processing of large data sets across clusters of computers using the map and reduce programming model. The architecture allows scaling up from a single server to thousands of machines. At the same time Hadoop delivers high-availability by detecting and handling failures at the application layer. The use of data replication guarantees the data reliability and fast access. The core Hadoop components are the Hadoop Distributed File System (HDFS) [9][10] and the MapReduce framework [11].

HDFS has a master/slave architecture with a *NameNode* as a master and multiple *DataNodes* as slaves. The *NameNode* is responsible for the storing and managing of all file structures, metadata, transactional operations and logs of the file system. The *DataNodes* store the actual data in the form of files. Each file is split into blocks of a preconfigured size. Every block is copied and stored on multiple *DataNodes*. The number of block copies depends on the *Replication Factor*.

MapReduce is a software framework that provides general programming interfaces for writing applications that process vast amounts of data in parallel, using a distributed file system, running on the cluster nodes. The MapReduce unit of work is called *job* and consists of input data and a MapReduce program. Each job is divided into *map* and *reduce* tasks. The map task takes a split, which is a part of the input data, and processes it according to the user-defined map function from the MapReduce program. The reduce task gathers the output data of the map tasks and merges them according to the user-defined reduce function. The number of reducers is specified by the user and does not depend on input splits or number of map tasks. The parallel application execution is achieved by running map tasks on each node to process the local data and then send the result to a reduce task which produces the final output.

Hadoop implements the MapReduce model by using two types of processes – *JobTracker* and *TaskTracker*. The *JobTracker* coordinates all jobs in Hadoop and schedules tasks to the *TaskTrackers* on every cluster node. The *TaskTracker* runs tasks assigned by the *JobTracker*.

Multiple other applications were developed on top of the Hadoop core components, also known as the Hadoop ecosystem, to make it more ease to use and applicable to variety of industries. Example for such applications are Hive [12], Pig [13], Mahout [14], HBase [15], Sqoop [16] and many more.

**YARN** (Yet Another Resource Negotiator) [17][18] is the next generation Apache Hadoop platform, which introduces new architecture by decoupling the programming model from the resource management infrastructure and delegating many scheduling-related functions to per-application components. This new design [18] offers some improvements over the older platform:
- Scalability
- Multi-tenancy
- Serviceability
- Locality awareness
- High Cluster Utilization
- Reliability/Availability
- Secure and auditable operation
- Support for programming model diversity
- Flexible Resource Model
- Backward compatibility

The major difference is that the functionality of the JobTracker is split into two new daemons – *ResourceManager* (RM) and *ApplicationMaster* (AM). The RM is a global service, managing all the resources and jobs in the platform. It consists of scheduler and *ApplicationManager*. The scheduler is responsible for allocation of resources to the various running applications based on their resource requirements. The *ApplicationManager* is responsible for accepting jobs-submissions and negotiating resources from the scheduler. The *NodeManager* (NM) agent runs on each worker. It is responsible for allocation and monitoring of node resources (CPU, memory, disk and network) usage and reports back to the *ResourceManager* (scheduler). An instance of the *ApplicationMaster* runs per-application on each node and negotiates the appropriate resource container from the scheduler. It is important to mention that the new MapReduce 2.0 maintains API compatibility with the older stable versions of Hadoop and therefore, MapReduce jobs can run unchanged.

**Cloudera Hadoop Distribution (CDH)** [19][20] is 100% Apache-licensed open source Hadoop distribution offered by Cloudera. It includes the core Apache Hadoop elements - Hadoop Distributed File System (HDFS) and MapReduce (YARN), as well as several additional projects from the Apache Hadoop Ecosystem. All components are tightly integrated to enable ease of use and managed by a central application - Cloudera Manager [21].

## 3. Experimental Setup

### 3.1. Hardware

The experiments were performed on a cluster consisting of 4 nodes connected directly through 1GBit Netgear switch, as shown on Figure 1. All 4 nodes are Dell PowerEdge T420 servers. The master node is equipped with 2x Intel Xeon E5-2420 (1.9GHz) CPUs each with 6 cores, 32GB of RAM and 1TB (SATA, 3.5 in, 7.2K RPM, 64MB Cache) hard drive. The worker nodes are equipped with 1x Intel Xeon E5-2420 (2.20GHz) CPU with 6 cores, 32GB of RAM and 4x 1TB (SATA, 3.5 in, 7.2K RPM, 64MB Cache) hard drives. More detailed specification of the node servers is provided in the Appendix (Table 12 and Table 13).

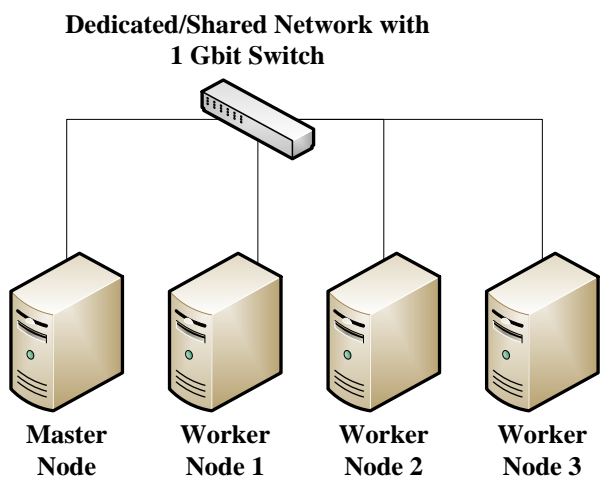

Figure 1: Cluster Setup

| Setup Description | Summary |
|---|---|
| *Total Nodes:* | 4 x Dell PowerEdge T420 |
| *Total Processors/ Cores/Threads :* | 5 CPUs/ 30 Cores/ 60 Threads |
| *Total Memory:* | 128 GB |
| *Total Number of Disks:* | 13 x 1TB, SATA, 3.5 in, 7.2K RPM, 64MB Cache |
| *Total Storage Capacity:* | 13 TB |
| *Network:* | 1 GBit Ethernet |

Table 1: Summary of Total System Resources

Table 1 summarizes the total cluster resources that are used in the calculation of the benchmark ratios in the next sections.

### 3.2. Software

This section describes the software setup of the cluster. The exact software versions that were used are listed in Table 2. Ubuntu Server LTS was installed on all 4 nodes, allocating the entire first disk. The number of open files per user was changed from the default value of 1024 to 65000 as suggested by the TPCx-HS benchmark and Cloudera guidelines [22]. Additionally, the OS *swappiness option* was turned permanently off (*vm.swappiness = 0*). The remaining three disks, on all worker nodes, were formatted as *ext4* partitions and permanently mounted with options *noatime* and *nodiratime*. Then the partitions were configured to be used by HDFS through the Cloudera Manager. Each 1TB disk provides in total 916.8GB of effective HDFS space, which means that all three workers (3 x 916.8GB = 8251.2GB = 8.0578TB) have in total around 8TB of effective HDFS space.

| Software | Version |
|---|---|
| *Ubuntu Server 64 Bit* | 14.04.1 LTS, Trusty Tahr, Linux 3.13.0-32-generic |
| *Java (TM) SE Runtime Environment* | 1.6.0_31-b04, 1.7.0_72-b14 |
| *Java HotSpot (TM) 64-Bit Server VM* | 20.6-b01, mixed mode 24.72-b04, mixed mode |
| *OpenJDK Runtime Environment* | 7u71-2.5.3-0ubuntu0.14.04.1 |
| *OpenJDK 64-Bit Server VM* | 24.65-b04, mixed mode |
| *Cloudera Hadoop Distribution* | 5.2.0-1.cdh5.2.0.p0.36 |
| *TPCx-HS Kit* | 1.1.2 |

Table 2: Software Stack of the System under Test

Cloudera CDH 5.2, with default configurations, was used for all experiments. Table 3 summarizes the software services running on each node. Due to the resource limitation (only 3 worker nodes) of our experimental setup, the cluster was configured to work with replication factor of 2. This means that our cluster can store at most 4TB of data.

| Server | Disk Drive | Software Services |
|---|---|---|
| *Master Node* | Disk 1/ sda1 | Operating System, Root, Swap, Cloudera Manager Services, Name Node, SecondaryName Node, Hive Metastore, Hive Server2, Oozie Server, Spark History Server, Sqoop 2 Server, YARN Job History Server, Resource Manager, Zookeeper Server |
| *Worker Nodes 1-3* | Disk 1/ sda1 | Operating System, Root, Swap, Data Node, YARN Node Manager |
| | Disk 2/ sdb1 | Data Node |
| | Disk 3/ sdc1 | Data Node |
| | Disk 4/ sdd1 | Data Node |

Table 3: Software Services per Node

### 3.3. Network Setups

The initial cluster setup was using the *shared 1GBit* network available in our lab. However, as expected it turned out that it does not provide sufficient network speed for network intensive cluster applications. Also it was hard to estimate the actual available bandwidth of the shared network as it consists of multiple workstation machine, which are utilizing it in a none predictable manner. Therefore, it was clear that a dedicated network should be setup for our cluster. This was achieved by using a simple 1GBit commodity switch (Netgear GS108 GE, 8-Port), which connected all four nodes directly in a *dedicated 1GBit network*.

To validate that our cluster setup was properly installed, the network speed was measured using a standard network tool called *iperf* [23]. Using the provided instructions [24][25], we obtained multiple measurements for our *dedicated 1GBit network* between the NameNode and two of our DataNodes, reported in Table 4. The iperf server was started by executing "$*iperf -s*" command on the NameNode and then executing two times the iperf clients using the "$*iperf -client* serverhostname *-time 30 -interval 5 -parallel 1 -dualtest*" command on the DataNodes. The iperf numbers show very stable data transfer of around 930 Mbits per second.

| Run | Server | Client | Time (sec) | Interval | Parallel | Type | Transfer 1 (Gbytes) | Speed 1 (Mbits/sec) | Transfer 2 (Gbytes) | Speed 2 (Mbits/sec) |
|---|---|---|---|---|---|---|---|---|---|---|
| 1 | NameNode | DataNode 1 | 30 | 5 | 1 | dualtest | 3.24 | 929 | 3.25 | 930 |
| 2 | NameNode | DataNode 1 | 30 | 5 | 1 | dualtest | 3.24 | 928 | 3.25 | 931 |
| 1 | NameNode | DataNode 2 | 30 | 5 | 1 | dualtest | 3.25 | 930 | 3.25 | 931 |
| 2 | NameNode | DataNode 2 | 30 | 5 | 1 | dualtest | 3.24 | 928 | 3.25 | 931 |

Table 4: Network Speed

The next steps is to test the different cluster setups using a network intensive Big Data benchmark like TPCx-HS in order to get a better idea of the implications of using a shared network versus a dedicated one.

## 4. Benchmarking Methodology

This section presents the TPCx-HS benchmark, its methodology and some of its major features as described in the current specification (version 1.3.0 from February 19, 2015) [1][2].
The TPCx-HS was released in July 2014 as the first industry's standard benchmark for Big Data systems. It stresses both the hardware and software components including the Hadoop run-time stack, Hadoop File System and MapReduce layers. The benchmark is based on the TeraSort workload [26], which is part of the Apache Hadoop distribution. Similarly, it consists of four modules: **HSGen**, **HSDataCkeck**, **HSSort** and **HSValidate**. The **HSGen** is a program that generates the data for a particular Scale Factor (see Clause 4.1 from the TPCx-HS specification) and is based on the TeraGen, which uses a random data generator. The **HSDataCheck** is a program that checks the compliance of the dataset and replication. **HSSort** is a program, based on TeraSort, which sorts the data into a total order. Finally, **HSValidate** is a program, based on TeraValidate, which validates if the output is correctly sorted.
A valid benchmark execution consists of five separate phases which has to be run sequentially to avoid any phase overlapping, as depicted on Figure 2. Additionally, Table 5 provides exact description of each of the execution phases. The benchmark is started by the <TPCx-HS-master> script and consists of two consecutive runs, Run1 and Run2, as shown on Figure 2. No activities except file system cleanup are allowed between Run1 and Run2. The completion times of each phase/module (HSGen, HSSort and HSValidate) except HSDataCheck are currently reported.
An important requirement of the benchmark is to maintain 3-way data replication throughout the entire experiment. In our case this criteria was not fulfilled, because of the limited resources of our clusters (only 3 worker nodes). All of our experiments were performed with 2-way data replication.
The benchmark reports the total elapsed time (T) in seconds for both runs. This time is used for the calculation of the TPCx-HS Performance Metric also abbreviated with HSph@SF. The run that takes more time and results in lower TPCx-HS Performance Metric is defined as the performance run. On the contrary, the run that takes less time and results in TPCx-HS Performance Metric is defined as the repeatability run. The benchmark reported performance metric is the TPCx-HS Performance Metric for the performance run.

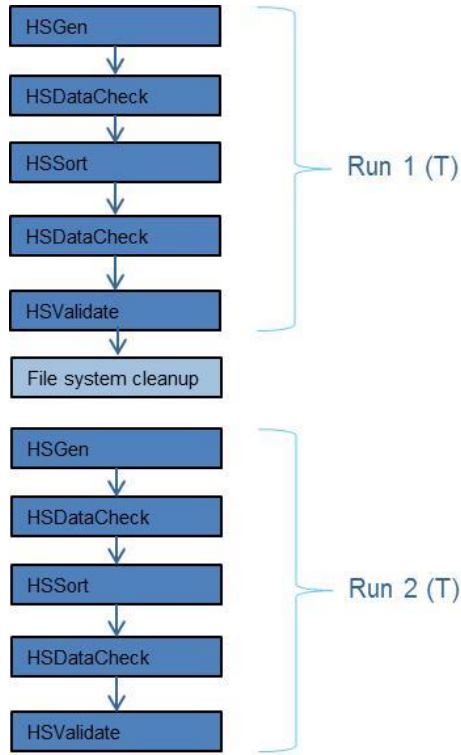

Figure 2: TPCx-HS Execution Phases (version 1.3.0 from February 19, 2015) [1]

| Phase | Description as provided in TPCx-HS specification (version 1.3.0 from February 19, 2015) [1] |
|---|---|
| 1 | *"Generation of input data via HSGen. The data generated must be replicated 3-ways and written on a Durable Medium."* |
| 2 | *"Dataset (See Clause 4) verification via HSDataCheck. The program is to verify the cardinality, size and replication factor of the generated data. If the HSDataCheck program reports failure then the run is considered invalid."* |
| 3 | *"Running the sort using HSSort on the input data. This phase samples the input data and sorts the data. The sorted data must be replicated 3-ways and written on a Durable Medium."* |
| 4 | *"Dataset (See Clause 4) verification via HSDataCheck. The program is to verify the cardinality, size and replication factor of the sorted data. If the HSDataCheck program reports failure then the run is considered invalid."* |
| 5 | *"Validating the sorted output data via HSValidate. HSValidate validates the sorted data. This phase is not part of the primary metric but reported in the Full Disclosure Report. If the HSValidate program reports that the HSSort did not generate the correct sort order, then the run is considered invalid. "* |

Table 5: TPCx-HS Phases

The *Scale Factor* defines the size of the dataset, which is generated by HSGen and used for the benchmark experiments. In TPCx-HS, it follows a stepped size model. Table 6 summarizes the supported *Scale Factors*, together with the corresponding data sizes and number of records. The last column indicates the argument with which to start the TPCx-HS-master script.

| Data Size | Scale Factor (SF) | Number of Records | Option to Start Run |
|---|---|---|---|
| 100 GB | 0.1 | 1 Billion | ./TPCx-HS-master.sh -g 1 |
| 300 GB | 0.3 | 3 Billions | ./TPCx-HS-master.sh -g 2 |
| 1 TB | 1 | 10 Billions | ./TPCx-HS-master.sh -g 3 |
| 3 TB | 3 | 30 Billions | ./TPCx-HS-master.sh -g 4 |
| 10 TB | 10 | 100 Billions | ./TPCx-HS-master.sh -g 5 |
| 30 TB | 30 | 300 Billions | ./TPCx-HS-master.sh -g 6 |
| 100 TB | 100 | 1000 Billions | ./TPCx-HS-master.sh -g 7 |
| 300 TB | 300 | 3000 Billions | ./TPCx-HS-master.sh -g 8 |
| 1 PB | 1000 | 10 000 Billions | ./TPCx-HS-master.sh -g 9 |

Table 6: TPCx-HS Scale Factors

The TPCx-HS specification defines three major metrics:

- *Performance metric (HSph@SF)*
- *Price-performance metric ($/HSph@SF)*
- *Power per performance metric* (*Watts/HSph@SF)*

The *performance metric (HSph@SF)* represents the effective sort throughput of the benchmark and is defined as:

$$HSph@SF = \frac{SF}{T/3600}$$

where **SF** is the *Scale Factor* (see Clause 4.1 from the TPCx-HS specification) and **T** is the total elapsed time in seconds for the *performance run*. 3600 seconds are equal to 1 hour.

The *price-performance metric ($/HSph@SF)* is defined and calculated as follows:

$$\$/HSph@SF = \frac{P}{HSph@SF}$$

where **P** is the total cost of ownership of the tested system. If the price is in currency other than US dollars, the units can be adjusted to the corresponding currency.

The last *power per performance metric*, which is not covered in our study, is expressed as *Watts/HSph@SF*, which have to be measured following the TPC-Energy requirements [27].

## 5. Experimental Results

This section presents the results of the performed experiments. The TPCx-HS benchmark was run with three scale factors 0.1, 0.3 and 1 which generate respectively 100GB, 300GB and 1TB datasets. These three runs were performed for a cluster setup using both shared and dedicated 1Gbit networks. In the first part are presented and evaluated the metrics reported by the TPCx-HS benchmark for the different experiments. In the second part is analyzed the utilization of cluster resources with respect to the two network setups.

### 5.1. System Ratios

The system ratios are additional metrics defined in the TPCx-HS specification to better describe the system under test. These are the *Data Storage Ratio* and the *Scale Factor to Memory Ratio*. Using the *Total Physical Storage* (13TB) and the *Total Memory* (128GB or 0.125TB) reported in Table 1, we calculate them as follows:

$$\boldsymbol{Data\ Storage\ Ratio} = \frac{Total\ Physical\ Storage}{Scale\ Factor}$$

$$\boldsymbol{Scale\ Factor\ to\ Memory\ Ratio} = \frac{Scale\ Factor}{Total\ Physical\ Memory}$$

Table 7 reports the two ratios for the three different scale factors used in the experiments.

| Scale Factor | Data Size | Data Storage Ratio | Scale Factor to Memory Ratio |
|---|---|---|---|
| 0.1 | 100 GB | 130 | 0.8 |
| 0.3 | 300 GB | 43.3 | 2.4 |
| 1 | 1 TB | 13 | 8 |

Table 7: TPCx-HS Related Ratios

### 5.2. Performance

In this section are evaluated the results of the multiple experiments. ***The presented results are obtained by executing the TPCx-HS kit provided on the official TPC website [1]. However, the reported times and metrics are experimental, not audited by any authorized organization and therefore not directly comparable with other officially published full disclosure reports.***
Figure 3 illustrates the times of the two cluster setups (shared and dedicated 1GBit networks) for the three datasets 100GB, 300GB and 1TB. It can be clearly observed that for all the cases the dedicated 1GBit setup performs around 5 times better than the shared setup. Similarly, Figure 4 shows that the dedicated setup **achieves between 5 and 6 times** more HSph@SF metric than the shared setup.

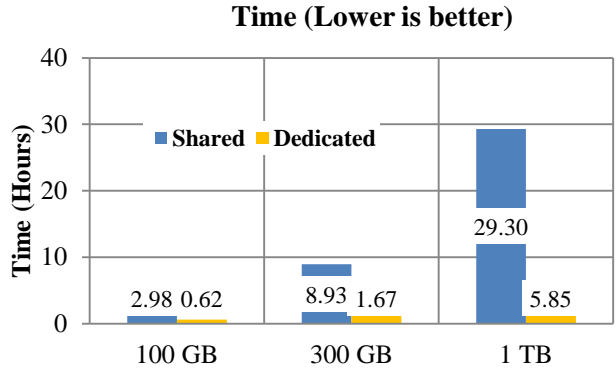
Figure 3: TPCx-HS Times

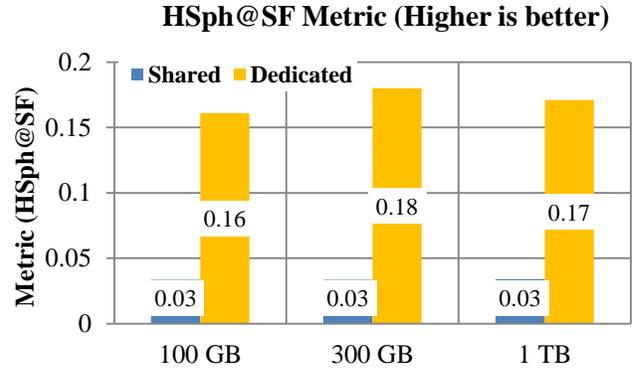
Figure 4: TPCx-HS Metric

Table 8 summarizes the experimental results introducing additional statistical comparisons. The **Data Δ** column represents the difference in percent of the Data Size to the data baseline in our case 100GB. In the first case, scale factor 0.3 increases the processed data with 200%, whereas in the second case the data is increased with 900%. The **Time (Sec)** shows the average time in seconds of two complete TPCx-HS runs for all the six test configurations. The following **Time Stdv (%)** shows the standard deviation of **Time (Sec)** in percent between the two runs. Finally, the **Time Δ (%)** represents the difference in percent of **Time (Sec)** to the time baseline in our case scale factor 0.1. Here we observe that for the shared setup the execution time takes five times longer than the dedicated setup.

| Scale Factor | Data Size | Data Δ (%) | Network | Metric (HSph@SF) | Time (Sec) | Time Stdv (%) | Time Δ (%) |
|---|---|---|---|---|---|---|---|
| 0.1 | 100 GB | baseline | shared | 0.03 | 10721.75 | 0.59 | baseline |
| 0.3 | 300 GB | +200 | shared | 0.03 | 32142.75 | 0.50 | +199.79 |
| 1 | 1 TB | +900 | shared | 0.03 | 105483.00 | 0.41 | +883.82 |
| | | | | | | | |
| 0.1 | 100 GB | baseline | dedicated | 0.16 | 2234.75 | 0.72 | baseline |
| 0.3 | 300 GB | +200 | dedicated | 0.18 | 6001.00 | 0.89 | +168.53 |
| 1 | 1 TB | +900 | dedicated | 0.17 | 21047.75 | 1.53 | +841.84 |

Table 8: TPCx-HS Results

Figure 5 illustrates the scaling behavior between the two network setups based on different data sizes. The results show that the dedicated setup has a better scaling behavior than the shared setup. Additionally, the increase of data size improves the scaling behavior of both setups.

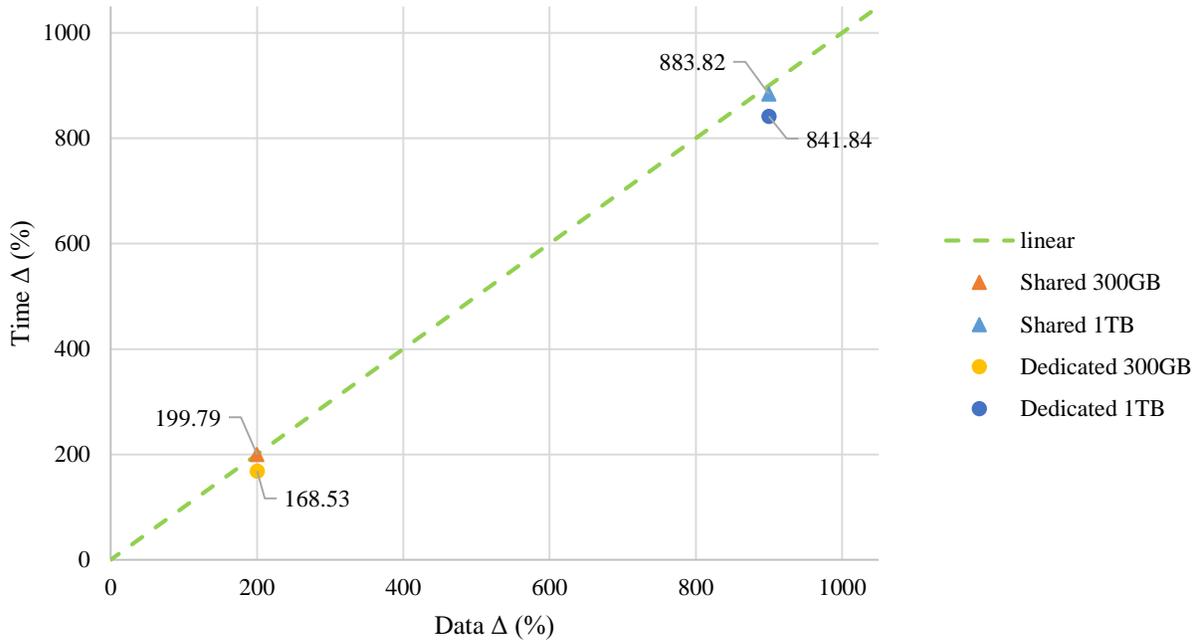

Figure 5: TPCx-HS Scaling Behavior (0 on the *X* and *Y-axis* is equal to the baseline of SF 0.1/100GB)

The TPCx-HS benchmark consists of five phases, which were explained in Section 4. Table 9 depicts the average times of the three major phases, together with their standard deviations in percent. Clearly the data sorting phase (HSSort) takes the most processing time, followed by the data generation phase (HSGen) and finally the data validation phase (HSValidate).

| Scale Factor/ Data Size | Network | HSGen (Sec) | HSGen Stdv (%) | HSSort (Sec) | HSSort Stdv (%) | HSValidate (Sec) | HSValidate Stdv (%) |
|---|---|---|---|---|---|---|---|
| 0.1 / 100 GB | shared | 3031.62 | 2.05 | 7391.83 | 1.18 | 289.02 | 0.90 |
| 0.3 / 300 GB | shared | 9149.36 | 0.99 | 22501.87 | 0.66 | 482.33 | 2.80 |
| 1 / 1 TB | shared | 29821.68 | 1.35 | 74432.82 | 1.21 | 1219.52 | 1.39 |
|  |  |  |  |  |  |  |  |
| 0.1 / 100 GB | dedicated | 543.06 | 0.47 | 1394.60 | 0.97 | 288.19 | 0.01 |
| 0.3 / 300 GB | dedicated | 1348.98 | 0.50 | 4112.33 | 0.57 | 530.97 | 7.27 |
| 1 / 1 TB | dedicated | 4273.30 | 0.16 | 15271.96 | 3.72 | 1493.41 | 6.35 |

Table 9: TPCx-HS Phase Times

## 5.3. Price-Performance

The price-performance metric of the TPCx-HS benchmark that reviewed in Section 4 divides the total cost of ownership (P) of the system under test on the TPCx-HS metric (HSph@SF) for a particular scale factor. The total cost of our cluster for the dedicated 1Gbit network setup is summarized in Table 10. It include only the hardware prices as the software that is used in the setup is available for free. There are no support, administrative and electricity costs included. The total price of the shared network setup is equal to 6730 €, which basically includes all the components listed in Table 10 without the network switch.

| Hardware Components | Price incl. Taxes (€) |
|---|---|
| 1 x Master Node (Dell PowerEdge T420) | 1803 |
| 1 x Switch (Netgear GS108 GE, 8-Port, 1 GBit) | 30 |
| 3 x Data Nodes (Dell PowerEdge T420) | 4228 |
| 9 x Disks (Western Digital Blue Desktop, 1TB) | 450 |
| Additional Costs (Cables, Mouse & Monitor) | 249 |
| Total Price (€) | 6760 |

Table 10: Total System Cost with 1GBit Switch

Table 11 summarizes the price-performance metric for the tested scale factors in the two network setups. *The lower price-performance metric indicates better system performance*. In our experiments this is the dedicated setup, which has **around 6 times smaller** price-performance than the shared setup.

| Scale Factor | Data Size | Network | Metric (HSph@SF) | Price-Performance Metric (€/HSph@SF) |
|---|---|---|---|---|
| 0.1 | 100 GB | shared | 0.03 | 224333.33 |
| 0.3 | 300 GB | shared | 0.03 | 224333.33 |
| 1 | 1 TB | shared | 0.03 | 224333.33 |
| | | | | |
| 0.1 | 100 GB | dedicated | 0.16 | 42250.00 |
| 0.3 | 300 GB | dedicated | 0.18 | 37555.56 |
| 1 | 1 TB | dedicated | 0.17 | 39764.71 |

Table 11: Price-Performance Metrics

Using the price-performance formula, we can also find the maximal P (maxP) and respectively the highest switch price for which the dedicated setup will still perform better than the shared setup.

$$\text{€/HSph@SF(shared)} = \frac{\text{P(shared)}}{\text{HSph@SF}} = \frac{6730}{0.03} = 224333.33$$

$$224333.33 > \frac{\text{maxP(dedicated)}}{0.18}$$

$$224333.33 \times 0.18 > \text{maxP(dedicated)}$$

$$40380.00 > \text{maxP(dedicated)}$$

The highest total system cost for the dedicated 1Gbit setup should be less than 40 380€ in order for the system to achieve better price-performance (€/HSph@SF) than the shared setup. This represents a difference of about 33 620€, which just shows how huge is the gain in terms of cost when adding the 1Gbit switch.
In summary, the dedicated 1Gbit network setup costs around 30 € more than the shared setup (due to the already existing network infrastructure), but it achieves *between 5 and 6 times* better performance.

## 6. Resource Utilization

The following section presents graphically the cluster resource utilization in terms of CPU, memory, network and number of map and reduce jobs. The reported statistics are obtained using the Performance Analysis Tool (PAT) [28] while running the TPCx-HS with scale factor 0.1 (100GB) for both shared and dedicated network setups. The average values obtained in the measurement are reported in Table 14 and Table 15 in the Appendix. The graphics represent complete benchmark run, consisting of Run1 and Run2 as described in Section 4, for both Master and Worker nodes. The goal is to compare and analyze the cluster resource utilization between the two different network setups.

### 6.1. CPU

Figure 6 shows the CPU utilization in percent of the Master node for the dedicated and shared network setups with respect to the elapsed time (in seconds). The **System %** (in green) represents the CPU utilization that occurred when executing at the kernel (system) level. Respectively, the **User %** (in red) represents the CPU utilization when executing at the application (user) level. Finally, the **IOwait %** (in blue) represent the time in which the CPU was idle waiting for an outstanding disk I/O request.
Comparing both graphs, one can observe a slightly higher CPU utilization in the case of dedicated 1Gbit network. However, in both cases the overall CPU utilization (System and User %) is around 2%, leaving the CPU heavily underutilized.

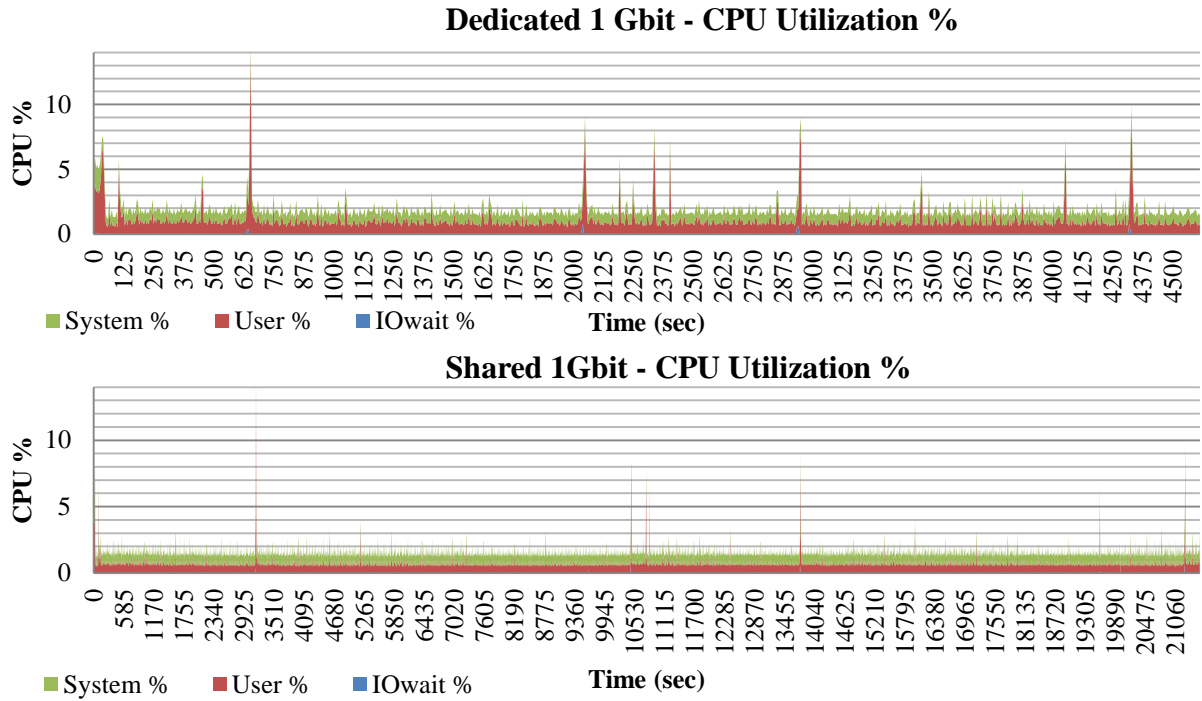

Figure 6: Master Node CPU Utilization

Similar to Figure 6, Figure 7 depicts the CPU utilization for one of the three Worker nodes. Clearly the dedicated setup utilizes better the CPU on both system and user level. In the same way, the IOwait times for the dedicated setup are much higher than the shared one. On average, the overall CPU utilization for the shared network is between 12% and 20%, whereas the overall for the dedicated network is between 56% and 70%. This difference is especially observed in the data generation and sorting phases, which are highly network intensive.

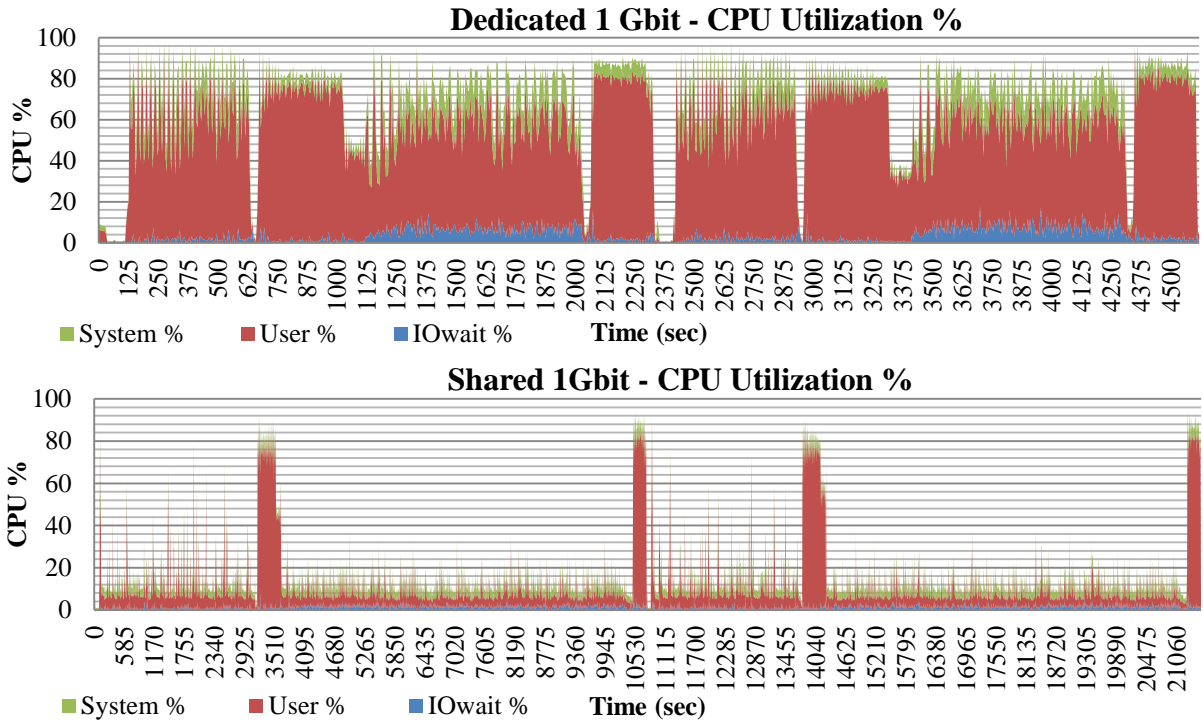

Figure 7: Worker Node CPU Utilization

Figure 8 illustrates the number of context switches per second, which measure the rate at which the threads/processes are switched in the CPU. The higher number of context switches indicates that the CPU spends more time on storing and restoring process states instead of doing real work [29]. In both graphics, we observe that the number of context switches per second is stable, on average between 10000 and 11000. This means that in both cases the Master node is equally utilized.

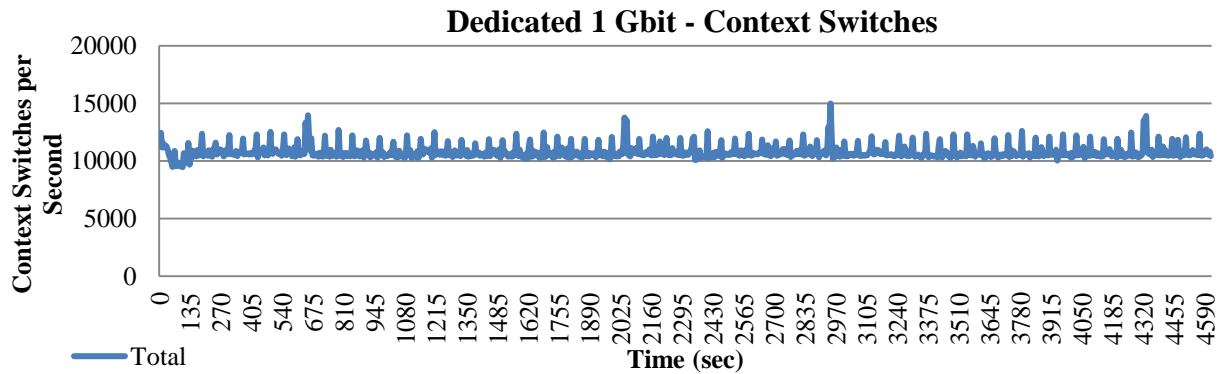

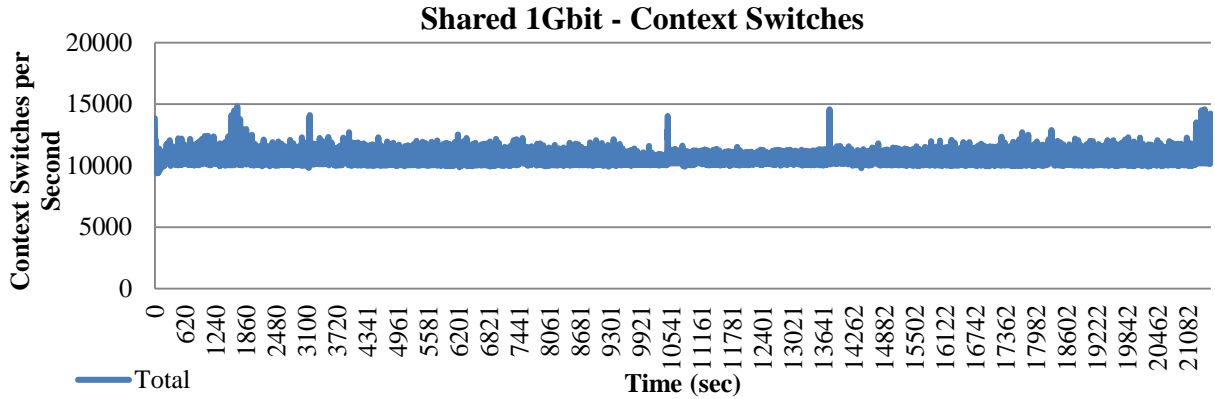

Figure 8: Master Node Context Switches

Similarly, Figure 9 depicts the context switches per second for one of the Worker nodes. In the dedicated 1Gbit case, we observe the number of context switches varies greatly in the different benchmark phases. The average number of context switches per second is around 20788. In the case of shared network, the average number of context switches per second is around 14233 and the variation between the phases is much smaller.

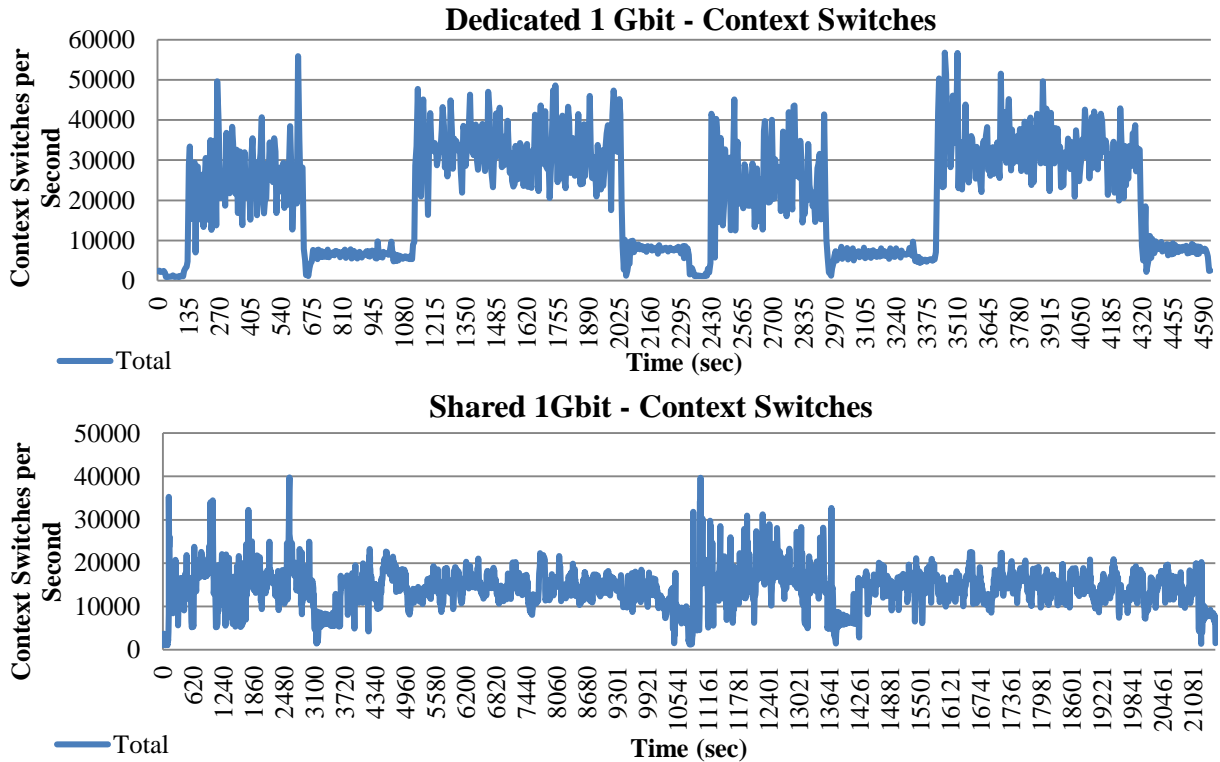

Figure 9: Worker Node Context Switches

## 6.2. Memory

Figure 10 shows the main memory utilization in percent of the Master node for the two network setups. In the dedicated 1Gbit setup the average memory used is around 48%, whereas in the shared setup it is around 91.4%.

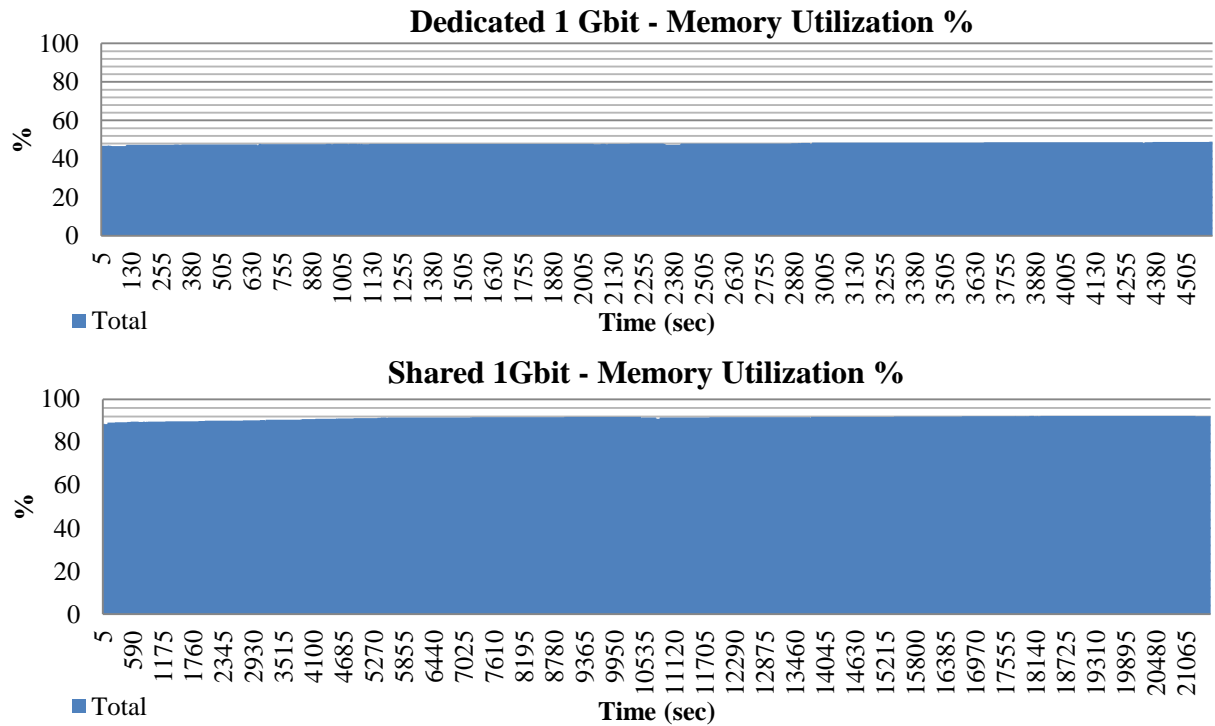

Figure 10: Master Node Memory Utilization

The same trend is observed on Figure 11, which depicts the free memory in Kbytes for the Master node. In the case of dedicated network, the average free memory is 16.3GB (17095562Kbytes), which is the remaining 52% of not utilized memory. Respectively, for the shared network the average free memory is around 2.7GB (2825635Kbytes), which is the remaining 8.6% of not utilized memory.

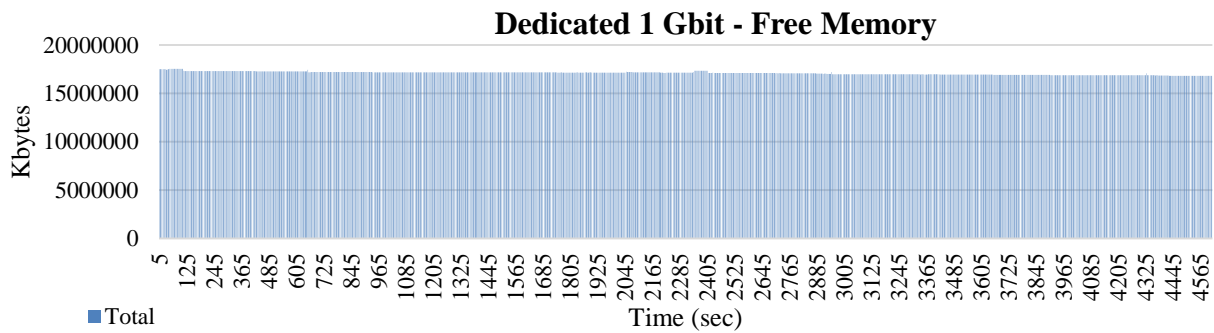

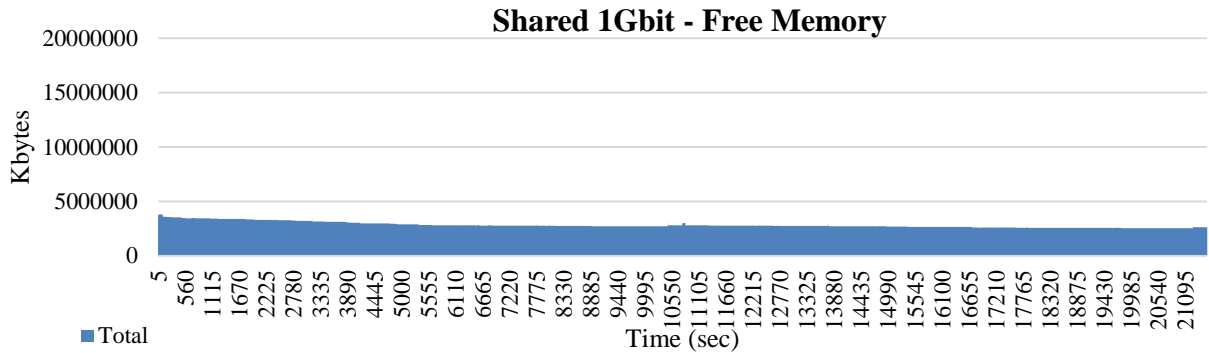

Figure 11: Master Node Free Memory

In the same way, Figure 12 illustrates the main memory utilization in percent for one of the Worker nodes. For the dedicated 1Gbit case, the average memory used is around 92.3%, whereas for the shared 1Gbit case it is around 92.9%. This confirms the great resemblance in both graphics, indicating that the Worker nodes are heavily utilized in the two setups. It will be advantageous to consider adding more memory to our nodes as it can further improve the performance by enabling more parallel jobs to be executed [29].

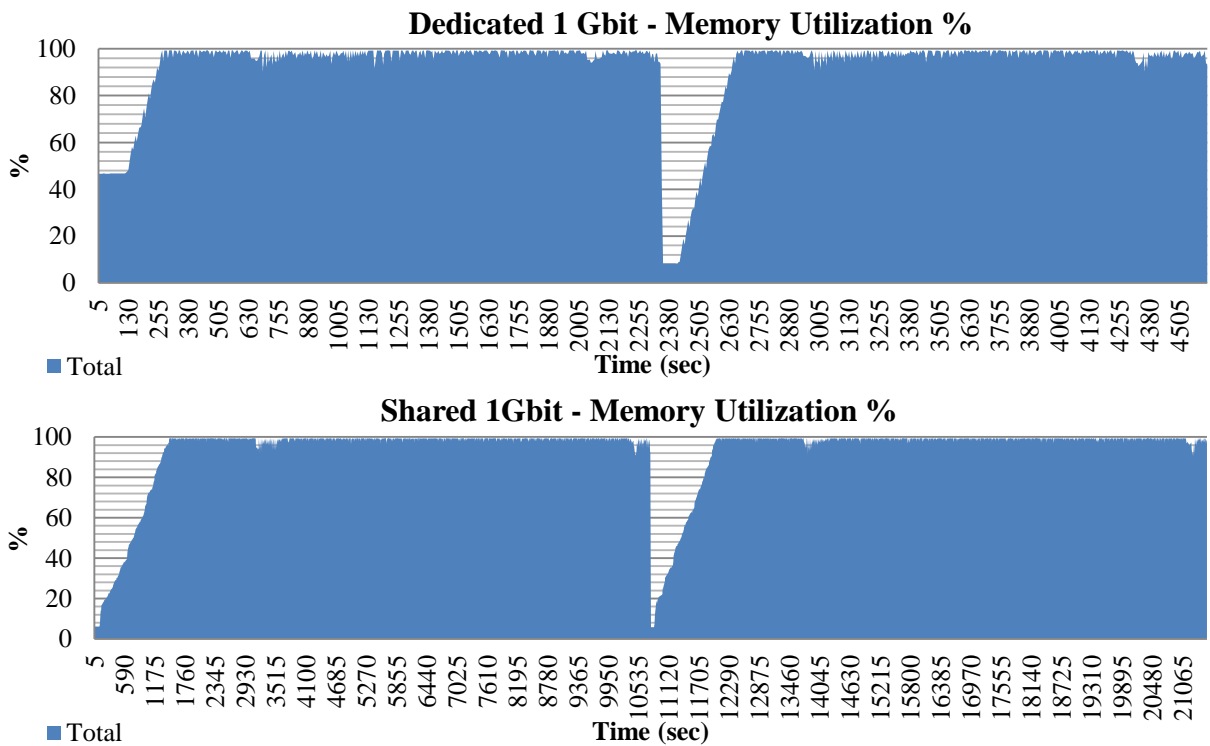

Figure 12: Worker Node Memory Utilization

Figure 13 shows the free memory and the amount of cached memory in Kbytes for one of the Worker nodes. For the dedicated 1Gbit setup, the average free memory is around 2.4GB (2528790Kbytes), which is exactly the 7.7% of non-utilized memory. Similarly, for the shared

1Gbit setup, the average free memory is around 2.2GB (2326411Kbytes) or around 7% not utilized memory.

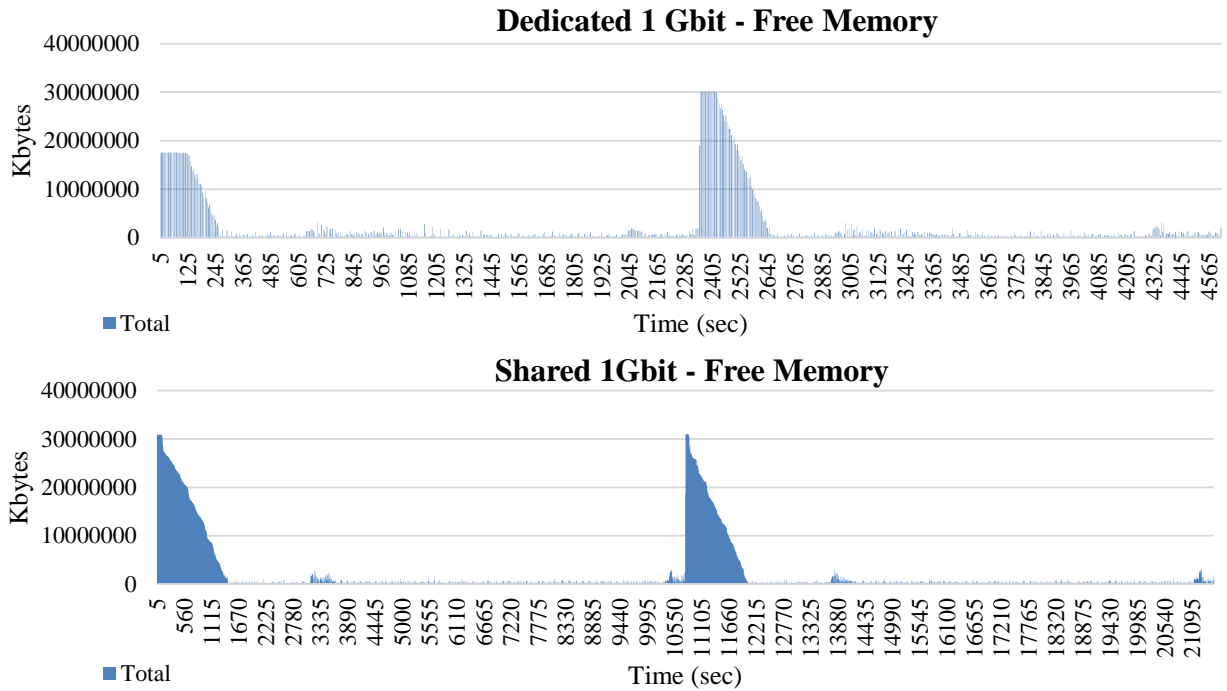

Figure 13: Worker Node Free Memory

## 6.3. Disk

The following graphics represent the number of read and write requests issued to the storage devices per second. Figure 14 shows that the Master node has very few read requests for the two network setups. On the other hand, the average number of write requests per second for the dedicated 1Gbit setup is around 3.1 and respectively around 1.7 for the shared network setup.

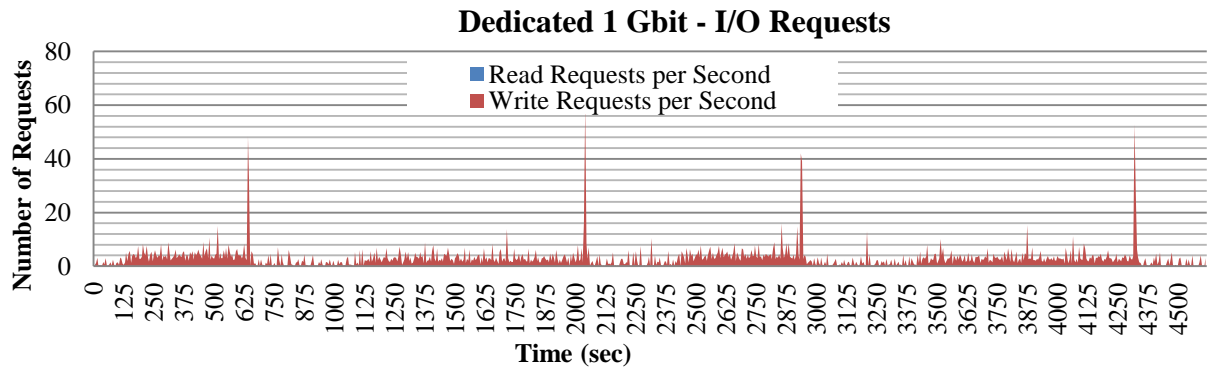

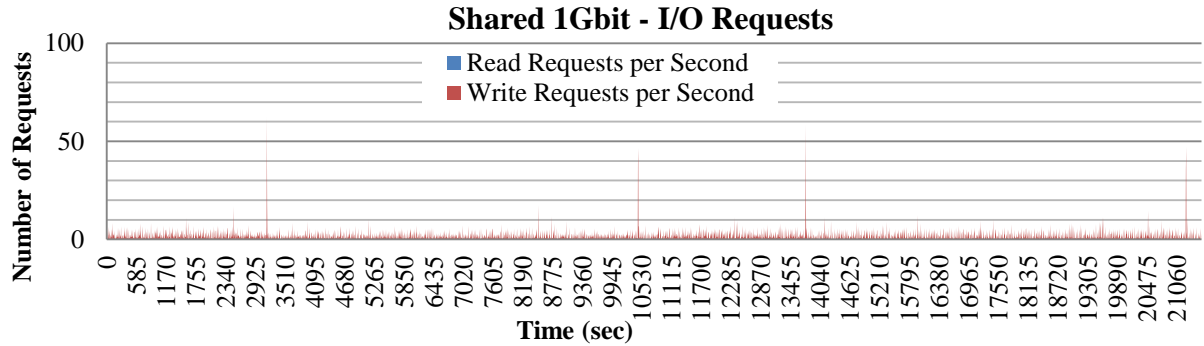

Figure 14: Master Node I/O Requests

Similarly, Figure 15 depicts the read and write requests per second for one of the Worker nodes. For the dedicated 1Gbit network setup, the average read requests per second are around 112 and respectively around 25 for the shared network setup. The average write requests per second for the dedicated network are around 40 and respectively around 10 for the shared network. This clearly indicates that the dedicated setup is much more efficient in reading and writing of data.

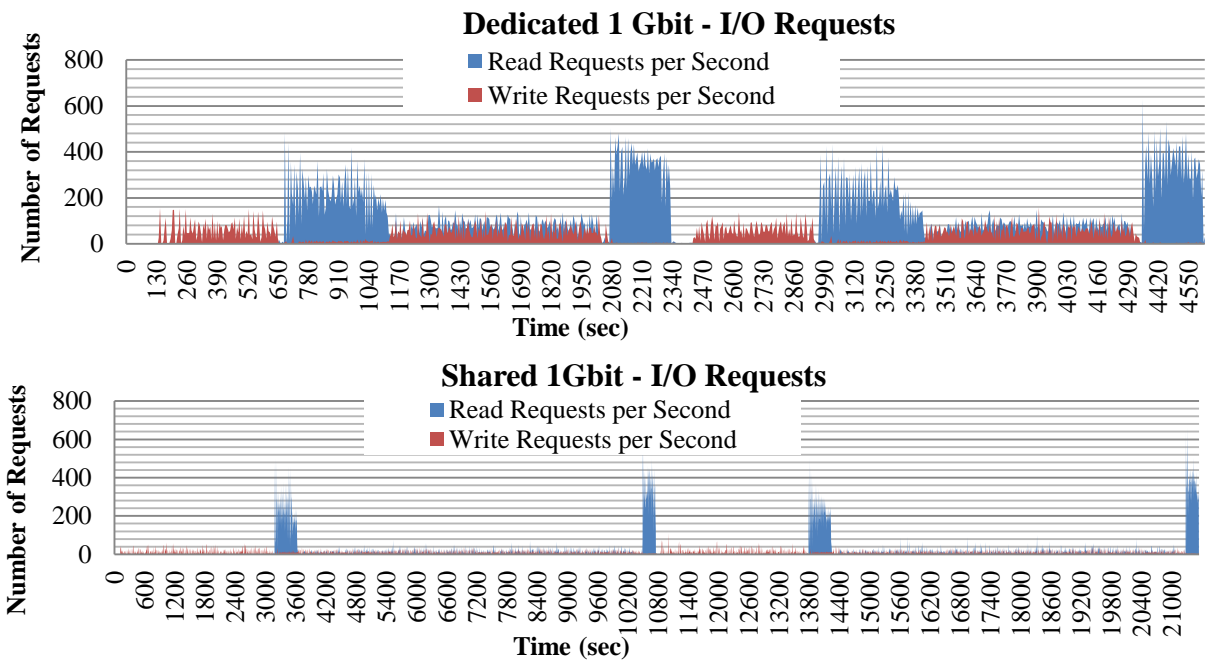

Figure 15: Worker Node I/O Requests

The following graphics illustrate the I/O latencies (in milliseconds), which is the average time spent from issuing of I/O requests to the time of their service by the device. This also includes the time spent in the device queue and the time for servicing them. Figure 16 shows the Master node latencies, which for the dedicated setup on average are around 0.15 milliseconds and respectively around 0.17 milliseconds for the shared setup.

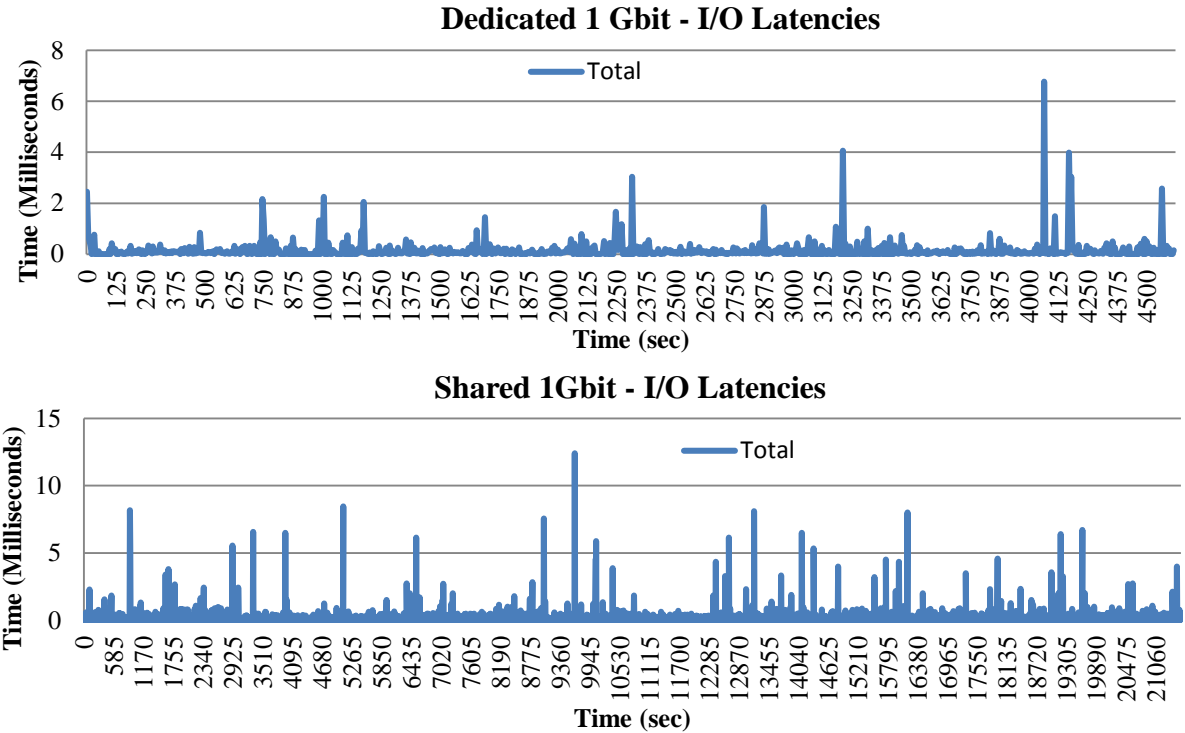

Figure 16: Master Node I/O Latencies

Similarly, Figure 17 depicts the Worker node latencies. For the dedicated setup, the average I/O latency is around 137 milliseconds and respectively around 70 milliseconds for the shared setup. We also observe that for the shared network setup, there are multiple I/O latencies taking around 200 to 300 milliseconds, whereas for the dedicated setup the longer latencies are much less.

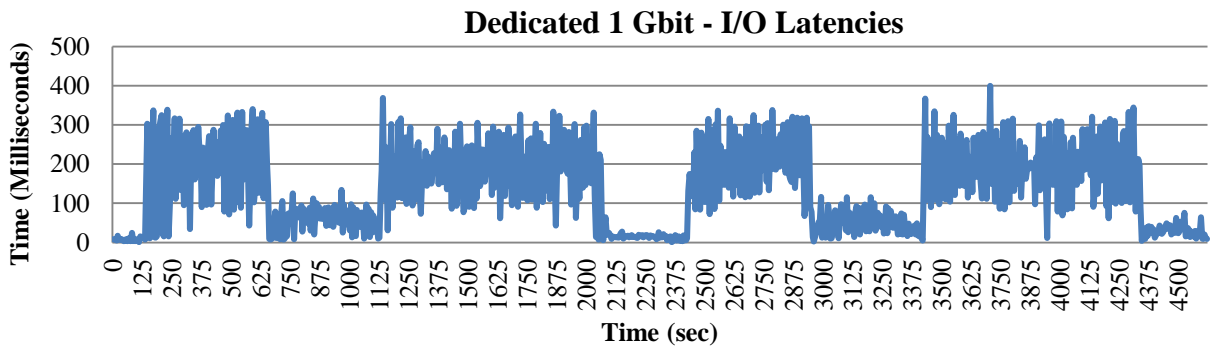

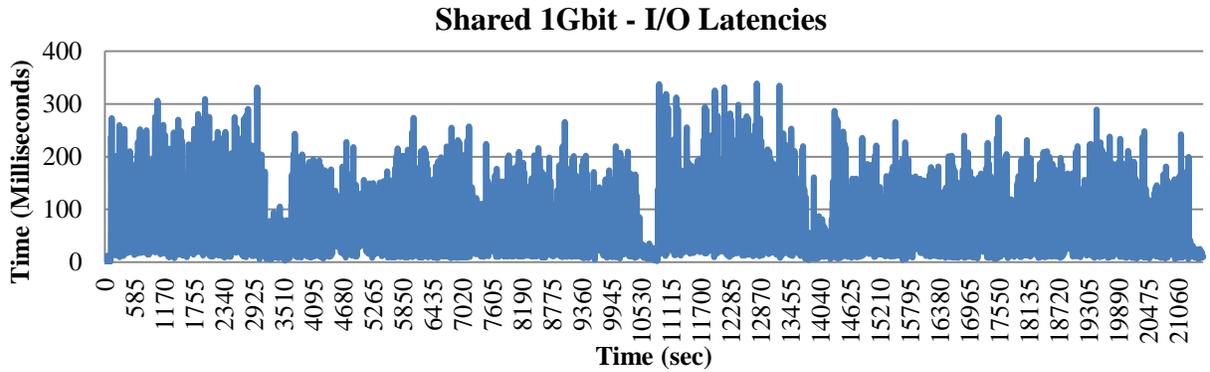

Figure 17: Worker Node I/O Latencies

The following figures depict the number of Kbytes read and written on the storage devices per second. Figure 18 illustrates this for the Master node. In both graphs, there are no read requests but only write one. On average around 31 Kbytes are written per second by the dedicated setup and respectively around 25 Kbytes are written per second by the shared setup.

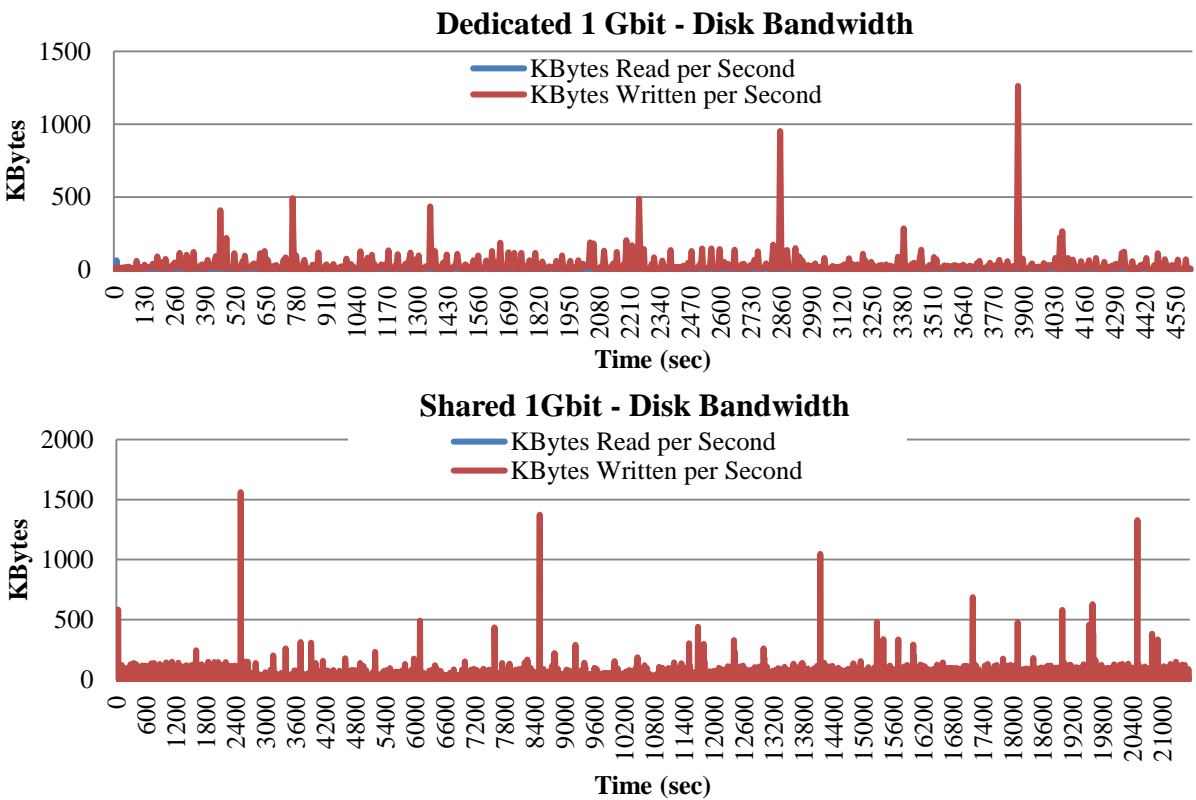

Figure 18: Master Node Disk Bandwidth

Similarly, Figure 19 shows the disk bandwidth for one of the Worker nodes. The average read throughput is around 6.4MB (6532Kbytes) per second for the dedicated setup and around 1.4MB (1438Kbytes) per second for the shared setup. Respectively the average write throughput for the dedicated setup is around 18.6MB (19010Kbytes) per second and 4MB (4087Kbytes) per second for the shared setup. In summary, the dedicated network achieves much better throughput levels, which indicates more efficient data processing and management.

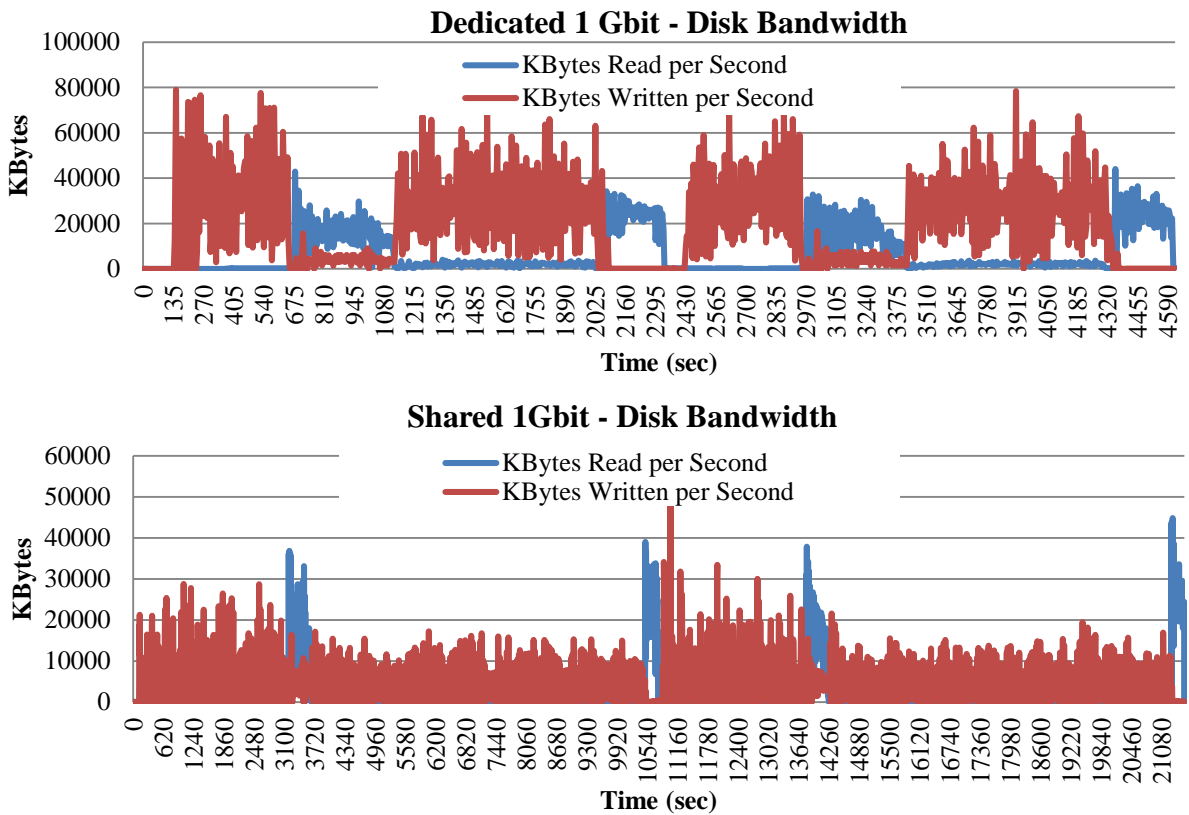

Figure 19: Worker Node Disk Bandwidth

### 6.4. Network

The following figures depict the number of received and transmitted Kbytes per second. For the Master node, the average number of received Kbytes per second is around 53 for the dedicated setups and around 22 for the shared setup. Respectively, the average transmitted Kbytes per second is around 42 for the dedicated configuration and around 19 for the shared configuration.

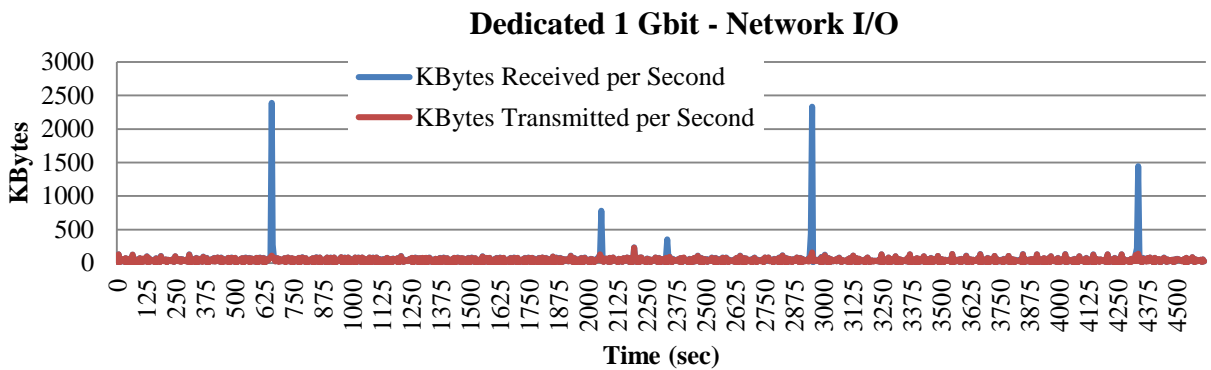

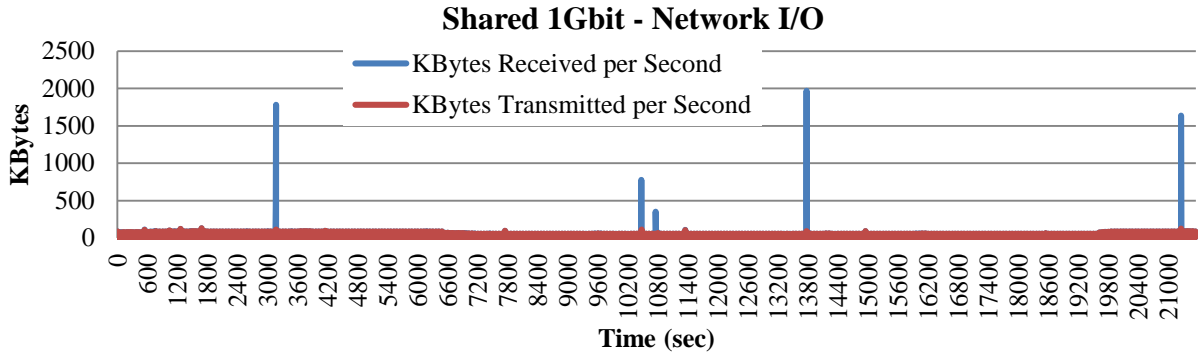

Figure 20: Master Node Network I/O

Analogously, Figure 21 shows the network transfer for one of the Worker nodes. In the case of dedicated setup, on average are received 32.8MB (33637Kbytes) per second and transmitted 30.6MB (31364Kbytes) per second. In the case of shared setup, on average are received 7.1MB (7297Kbytes) per second and transmitted 6.4MB (6548Kbytes) per second. Obviously, the dedicated 1Gbit network achieves **almost 5 times better utilization of the network**, resulting in faster overall performance.

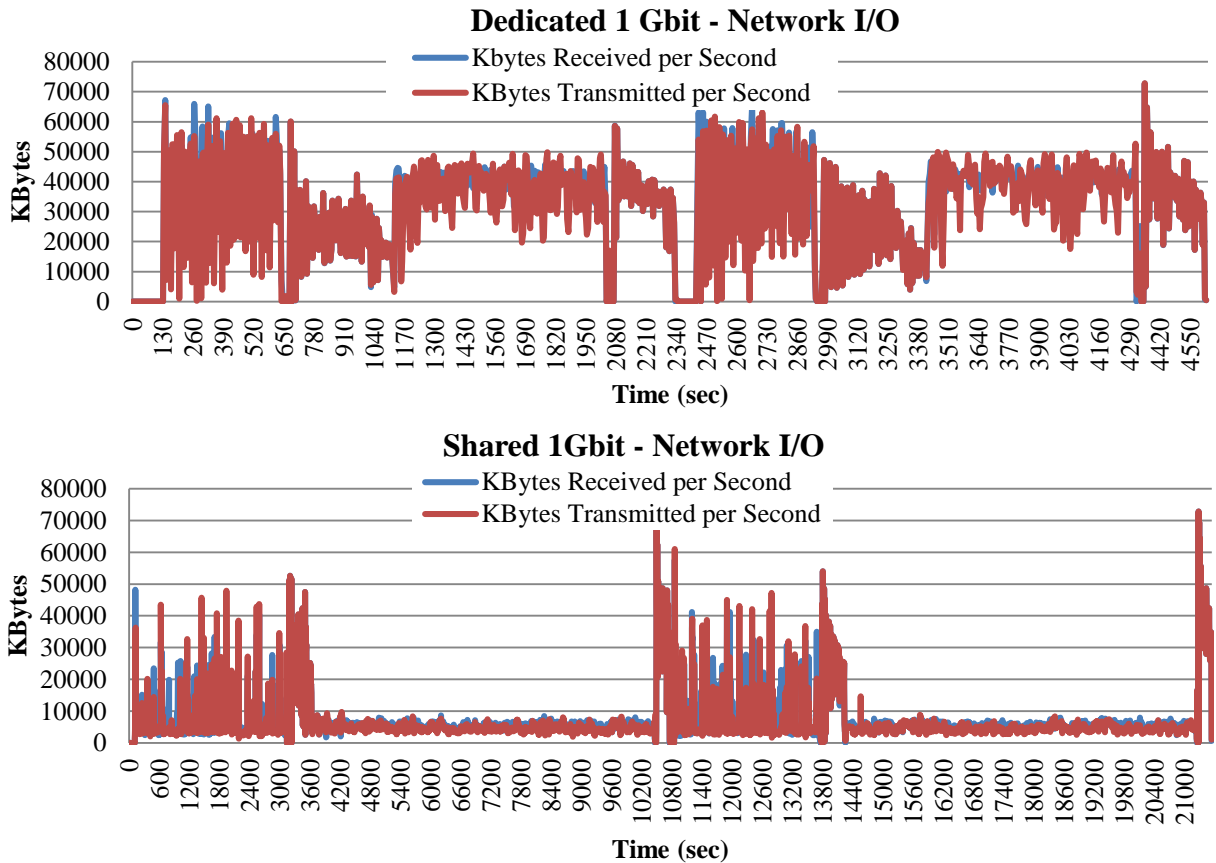

Figure 21: Worker Node Network I/O

## 6.5. Mappers and Reducers

Figure 22 shows the number of active map and reduce jobs in the different benchmark phases for both network setups. The behavior of the two graphics is very similar, except the fact that the shared setup is around 5 times slower than the dedicated setup.

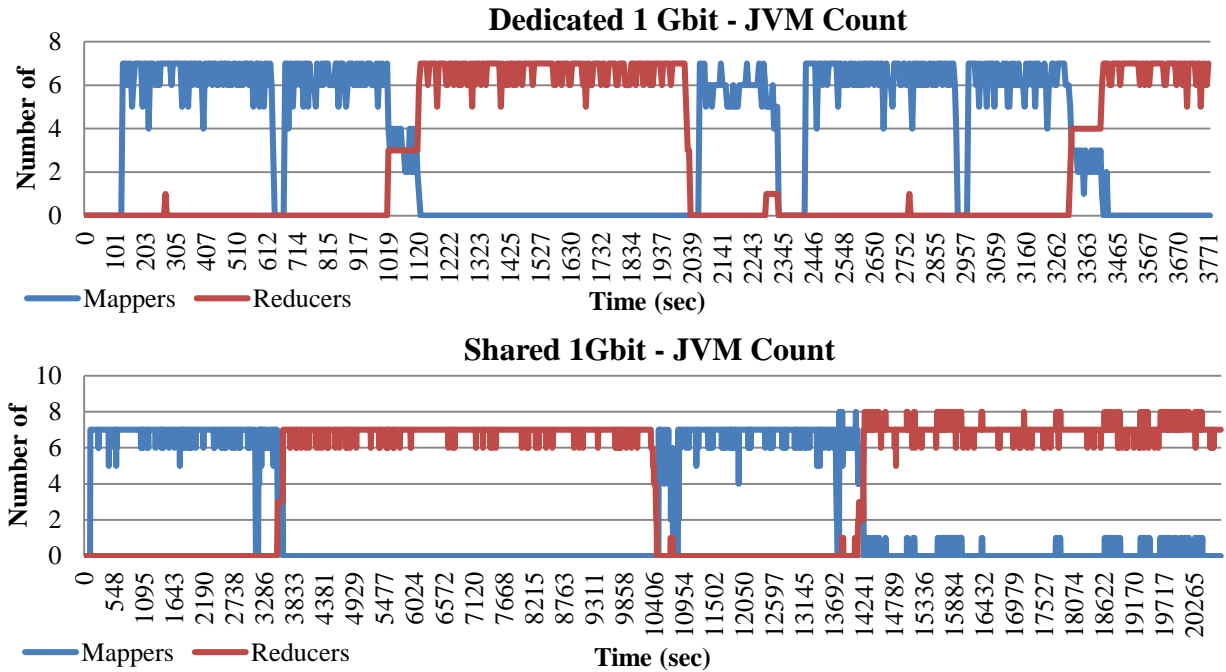

Figure 22: Worker Node JVM Count

## 7. Lessons Learned

The report presents a performance evaluation of the Cloudera Hadoop Distribution through the use of the TPCx-HS benchmark, which is the first officially standardized Big Data benchmark. In particular, our experiments compare two cluster setups: the first one using shared 1Gbit network and the second one using dedicated 1Gbit network. Our results show that the cluster with dedicated network setup is around *5 times faster* than the cluster with shared network setup in terms of:
- Execution time
- HSph@SF metric
- Average read and write throughput per second
- Network utilization

On average, the overall CPU utilization for the shared case is **between 12% and 20%**, whereas for the dedicated network it is **between 56% and 70%.** The average main memory usage, for the dedicated setup is *around 92.3%*, whereas for the shared setup it is *around 92.9%*.

Furthermore, based on the price-performance formula it can be concluded that the 5 times performance gain for the dedicated 1Gbit setup is equal to around 33 620€ in terms of money when compared to the shared 1Gbit setup. In the future, we plan to stress test the dedicated setup with other widely used Big Data benchmarks and workloads.

# References

[1] TPC, "TPCx-HS." [Online]. Available: http://www.tpc.org/tpcx-hs/.

[2] R. Nambiar, M. Poess, A. Dey, P. Cao, T. Magdon-Ismail, D. Q. Ren, and A. Bond, "Introducing TPCx-HS: The First Industry Standard for Benchmarking Big Data Systems," in *Performance Characterization and Benchmarking. Traditional to Big Data*, R. Nambiar and M. Poess, Eds. Springer International Publishing, 2014, pp. 1–12.

[3] J. Manyika, M. Chui, B. Brown, J. Bughin, R. Dobbs, C. Roxburgh, and A. H. Byers, "Big data: The next frontier for innovation, competition, and productivity," *McKinsey Glob. Inst.*, pp. 1–137, 2011.

[4] H. V. Jagadish, J. Gehrke, A. Labrinidis, Y. Papakonstantinou, J. M. Patel, R. Ramakrishnan, and C. Shahabi, "Big data and its technical challenges," *Commun. ACM*, vol. 57, no. 7, pp. 86–94, 2014.

[5] H. Hu, Y. Wen, T. Chua, and X. Li, "Towards Scalable Systems for Big Data Analytics: A Technology Tutorial," 2014.

[6] T. Ivanov, N. Korfiatis, and R. V. Zicari, "On the inequality of the 3V's of Big Data Architectural Paradigms: A case for heterogeneity," *ArXiv Prepr. ArXiv13110805*, 2013.

[7] R. Cattell, "Scalable SQL and NoSQL data stores," *ACM SIGMOD Rec.*, vol. 39, no. 4, pp. 12–27, 2011.

[8] "Apache Hadoop Project." [Online]. Available: http://hadoop.apache.org/. [Accessed: 22-May-2014].

[9] D. Borthakur, "The hadoop distributed file system: Architecture and design," *Hadoop Proj. Website*, vol. 11, p. 21, 2007.

[10] K. Shvachko, H. Kuang, S. Radia, and R. Chansler, "The hadoop distributed file system," in *Mass Storage Systems and Technologies (MSST), 2010 IEEE 26th Symposium on*, 2010, pp. 1–10.

[11] J. Dean and S. Ghemawat, "MapReduce: simplified data processing on large clusters," *Commun. ACM*, vol. 51, no. 1, pp. 107–113, 2008.

[12] A. Thusoo, J. S. Sarma, N. Jain, Z. Shao, P. Chakka, N. Zhang, S. Antony, H. Liu, and R. Murthy, "Hive - a petabyte scale data warehouse using Hadoop," in *2010 IEEE 26th International Conference on Data Engineering (ICDE)*, 2010, pp. 996–1005.

[13] A. F. Gates, O. Natkovich, S. Chopra, P. Kamath, S. M. Narayanamurthy, C. Olston, B. Reed, S. Srinivasan, and U. Srivastava, "Building a high-level dataflow system on top of Map-Reduce: the Pig experience," *Proc. VLDB Endow.*, vol. 2, no. 2, pp. 1414–1425, 2009.

[14] "Apache Mahout: Scalable machine learning and data mining." [Online]. Available: http://mahout.apache.org/. [Accessed: 30-Oct-2013].

[15] L. George, *HBase: the definitive guide*. O'Reilly Media, Inc., 2011.

[16] K. Ting and J. J. Cecho, *Apache Sqoop Cookbook*. O'Reilly Media, 2013.

[17] Apache YARN, "YARN," *Apache Hadoop NextGen MapReduce*, 2014. [Online]. Available: http://hadoop.apache.org/docs/current2/hadoop-yarn/hadoop-yarn-site/YARN.html. [Accessed: 06-Feb-2014].

[18] V. K. Vavilapalli, A. C. Murthy, C. Douglas, S. Agarwal, M. Konar, R. Evans, T. Graves, J. Lowe, H. Shah, S. Seth, B. Saha, C. Curino, O. O'Malley, S. Radia, B. Reed, and E. Baldeschwieler, "Apache Hadoop YARN: Yet Another Resource Negotiator," in *Proceedings of the 4th Annual Symposium on Cloud Computing*, New York, NY, USA, 2013, pp. 5:1–5:16.

[19]   Cloudera, "CDH," 2014. [Online]. Available: http://www.cloudera.com/content/cloudera/en/products-and-services/cdh.html. [Accessed: 28-Nov-2014].

[20]   Cloudera, "CDH Datasheet," 2014. [Online]. Available: http://www.cloudera.com/content/cloudera/en/resources/library/datasheet/cdh-datasheet.html. [Accessed: 28-Nov-2014].

[21]   Cloudera, "Cloudera Manager Datasheet," 2014. [Online]. Available: http://www.cloudera.com/content/cloudera/en/resources/library/datasheet/cloudera-manager-4-datasheet.html.

[22]   Cloudera, "Configuration Parameters: What can you just ignore?" [Online]. Available: http://blog.cloudera.com/blog/2009/03/configuration-parameters-what-can-you-just-ignore/. [Accessed: 11-Nov-2014].

[23]   "iPerf - The TCP, UDP and SCTP network bandwidth measurement tool." [Online]. Available: https://iperf.fr/. [Accessed: 04-Oct-2015].

[24]   "Measuring Network Performance: Test Network Throughput, Delay-Latency, Jitter, Transfer Speeds, Packet loss & Reliability. Packet Generation Using Iperf / Jperf," *Firewall.cx*. [Online]. Available: http://www.firewall.cx/networking-topics/general-networking/970-network-performance-testing.html. [Accessed: 04-Oct-2015].

[25]   J. Poelman, "IBM InfoSphere BigInsights Best Practices : Validating performance of a new cluster," 10-Jul-2014. [Online]. Available: https://www.ibm.com/developerworks/community/wikis/home?lang=en#!/wiki/W265aa64a4f21_43ee_b236_c42a1c875961/page/Validating%20performance%20of%20a%20new%20cluster. [Accessed: 04-Oct-2015].

[26]   Apache Hadoop, "TeraSort." [Online]. Available: http://hadoop.apache.org/docs/current/api/org/apache/hadoop/examples/terasort/package-summary.html. [Accessed: 14-Oct-2013].

[27]   Transaction Processing Performance Council, "TPC-Energy - Homepage." [Online]. Available: http://www.tpc.org/tpc_energy/default.asp. [Accessed: 15-Jul-2015].

[28]   Intel, "PAT at GitHub," 2014. [Online]. Available: https://github.com/intel-hadoop/PAT/tree/master/PAT. [Accessed: 09-Feb-2015].

[29]   K. Tannir, *Optimizing Hadoop for MapReduce*. Packt Publishing Ltd, 2014.

# Appendix

| System Information | | Description |
|---|---|---|
| Manufacturer: | Dell Inc. | |
| Product Name: | PowerEdge T420 | |
| BIOS: | 1.5.1 | Release Date: 03/08/2013 |
| Memory | | |
| Total Memory: | 32 GB | |
| DIMMs: | 10 | |
| Configured Clock Speed: | 1333 MHz | Part Number: M393B5273CH0-YH9 |
| Size: | 4096 MB | |
| CPU | | |
| Model Name: | Intel(R) Xeon(R) CPU E5-2420 0 @ 1.90GHz | |
| Architecture: | x86_64 | |
| CPU(s): | 24 | |
| On-line CPU(s) list: | 0-23 | |
| Thread(s) per core: | 2 | |
| Core(s) per socket: | 6 | |
| Socket(s): | 2 | |
| CPU MHz: | 1200.000 | |
| L1d cache: | 32K | |
| L1i cache: | 32K | |
| L2 cache: | 256K | |
| L3 cache: | 15360K | |
| NUMA node0 CPU(s): | 0,2,4,6,8,10,12,14,16,18,20,22 | |
| NUMA node1 CPU(s): | 1,3,5,7,9,11,13,15,17,19,21,23 | |
| NIC | | |
| Settings for em1: | Speed: 1000Mb/s | |
| Ethernet controller: | Broadcom Corporation NetXtreme BCM5720 Gigabit Ethernet PCIe | |
| Storage | | |
| Storage Controller: | LSI Logic / Symbios Logic MegaRAID SAS 2008 [Falcon] (rev 03) | 08:00.0 RAID bus controller |
| Drive / Name | Usable Space | Model |
| Disk 1/ sda1 | 931.5 GB | Western Digital, WD1003FBYX RE4-1TB, SATA3, 3.5 in, 7200RPM, 64MB Cache |

Table 12: Master Node Specifications

| System Information | | Description |
|---|---|---|
| Manufacturer: | Dell Inc. | |
| Product Name: | PowerEdge T420 | |
| BIOS: | 2.1.2 | Release Date: 01/20/2014 |
| Memory | | |
| Total Memory: | 32 GB | |
| DIMMs: | 4 | |
| Configured Clock Speed: | 1600 MHz | Part Number: M393B2G70DB0-YK0 |
| Size: | 16384 MB | |
| CPU | | |
| Model Name: | Intel(R) Xeon(R) CPU E5-2420 v2 @ 2.20GHz | |
| Architecture: | x86_64 | |
| CPU(s): | 12 | |
| On-line CPU(s) list: | 0-11 | |
| Thread(s) per core: | 2 | |
| Core(s) per socket: | 6 | |
| Socket(s): | 1 | |
| CPU MHz: | 2200.000 | |
| L1d cache: | 32K | |
| L1i cache: | 32K | |
| L2 cache: | 256K | |
| L3 cache: | 15360K | |
| NUMA node0 CPU(s): | 0-11 | |
| NIC | | |
| Settings for em1: | Speed: 1000Mb/s | |
| Ethernet controller: | Broadcom Corporation NetXtreme BCM5720 Gigabit Ethernet PCIe | |
| Storage | | |
| Storage Controller: | Intel Corporation C600/X79 series chipset SATA RAID Controller (rev 05) | 00:1f.2 RAID bus controller |
| Drive / Name | Usable Space | Model |
| Disk 1/ sda1 | 931.5 GB | Dell- 1TB, SATA3, 3.5 in, 7200RPM, 64MB Cache |
| Disk 2/ sdb1 | 931.5 GB | WD Blue Desktop WD10EZEX - 1TB, SATA3, 3.5 in, 7200RPM, 64MB Cache |
| Disk 3/ sdc1 | 931.5 GB | |
| Disk 4/ sdd1 | 931.5 GB | |

Table 13: Data Node Specifications

| Master Node | | |
|---|---|---|
| Network Type: | Dedicated 1Gbit | Shared 1Gbit |
| Scale Factor: | 100GB | 100GB |
| Avg. CPU Utilization % - User % | 1.09 | 0.75 |
| Avg. CPU Utilization % - System % | 0.83 | 0.78 |
| Avg. CPU Utilization % - IOwait % | 0.02 | 0.01 |
| Memory Utilization % | 48.04 | 91.41 |
| Avg. Kbytes Transmitted per Second | 42.28 | 18.96 |
| Avg. Kbytes Received per Second | 52.66 | 21.78 |
| Avg. Context Switches per Second | 10756.60 | 10360.07 |
| Avg. Kbytes Read per Second | 0.14 | 0.00 |
| Avg. Kbytes Written per Second | 31.39 | 25.23 |
| Avg. Read Requests per Second | 0.01 | 0.00 |
| Avg. Write Requests per Second | 3.12 | 1.74 |
| Avg. I/O Latencies in Milliseconds | 0.15 | 0.17 |

Table 14: Master Node - Resource Utilization

| Worker Node | | |
|---|---|---|
| Network Type: | Dedicated 1Gbit | Shared 1Gbit |
| Scale Factor: | 100GB | 100GB |
| Avg. CPU Utilization % - User % | 56.04 | 12.44 |
| Avg. CPU Utilization % - System % | 9.52 | 3.71 |
| Avg. CPU Utilization % - IOwait % | 3.61 | 1.28 |
| Memory Utilization % | 92.31 | 92.93 |
| Avg. Kbytes Transmitted per Second | 31363.57 | 6548.16 |
| Avg. Kbytes Received per Second | 33636.93 | 7297.27 |
| Avg. Context Switches per Second | 20788.16 | 14233.08 |
| Avg. Kbytes Read per Second | 6532.24 | 1438.23 |
| Avg. Kbytes Written per Second | 19010.01 | 4087.33 |
| Avg. Read Requests per Second | 111.75 | 24.70 |
| Avg. Write Requests per Second | 39.50 | 9.71 |
| Avg. I/O Latencies in Milliseconds | 136.87 | 69.83 |

Table 15: Worker Node - Resource Utilization

## Acknowledgements

This research is supported by the Frankfurt Big Data Lab at the Chair for Databases and Information Systems (DBIS) at the Goethe University Frankfurt. The authors would like to thank Naveed Mushtaq, Karsten Tolle, Marten Rosselli and Roberto V. Zicari for their helpful comments and support. We would like to thank Jeffrey Buell of VMware for his valuable feedback and corrections.